\documentclass[sigconf]{aamas}  
\usepackage{booktabs}

\setcopyright{ifaamas}  
\acmDOI{}  
\acmISBN{}  
\acmConference[AAMAS'19]{Proc.\@ of the 18th International Conference on Autonomous Agents and Multiagent Systems (AAMAS 2019)}{May 13--17, 2019}{Montreal, Canada}{N.~Agmon, M.~E.~Taylor, E.~Elkind, M.~Veloso (eds.)}  
\acmYear{2019}  
\copyrightyear{2019}  
\acmPrice{}  

\usepackage{balance}  

\renewcommand{\vec}[1]{\boldsymbol{#1}}

\DeclareMathOperator*{\argmax}{arg\,max}
\def\truth{\texttt{Truth}}
\def\KP{\texttt{KP}}

\def\AT{\texttt{AT}}
\def\CV{\texttt{CV}}
\def\LD{\texttt{LD}}
\def\LDLB{\texttt{LDLB}}

\def\NN{\texttt{NN}}

\def\AVU{\texttt{AU}}
\def\fsp{\vspace{-3mm}}

\usepackage{color}
\usepackage{graphicx}
\def\cent{\text{\textcent}}

\newcount\Comments  
\definecolor{darkgreen}{rgb}{0,0.6,0}
\newcommand{\kibitz}[2]{\ifnum\Comments=1{\textcolor{#1}{#2}}\fi}
\newcommand{\rmr}[1]{\kibitz{blue}{[RM:#1]}}
\newcommand{\kg}[1]{\kibitz{red}{[KG:#1]}}
\newcommand{\rf}[1]{\kibitz{green}{[RF:#1]}}

\newcommand{\newpar}[1]{\vspace{2mm}\noindent\textbf{#1.}}
\newcommand{\newparc}[1]{\vspace{2mm}\noindent\textbf{#1:}}
\newcommand{\newparn}[1]{\vspace{2mm}\noindent\textbf{#1}}

\begin{document}

\title{Modeling  People's   Voting Behavior with Poll Information}
  \author{Roy Fairstein and Adam Lauz}  
    \affiliation{
         \institution{Ben-Gurion Univ. of the Negev, Israel}
        }

\author{Reshef Meir}
    \affiliation{
         \institution{Technion, Israel Institute of Technology} 
        }
\author{Kobi Gal}  
    \affiliation{
         \institution{Ben-Gurion Univ. of the Negev, Israel, University of Edinburgh, U.K.}
        }



\begin{abstract}
 Despite the prevalence of voting systems in the real world
 there is no consensus among researchers  of how people vote strategically, even in simple   voting settings.
   This paper addresses this gap by comparing different   approaches that have been used to model  strategic voting,
  including  expected utility maximization, heuristic decision-making, and bounded
   rationality models. 
     The models  are applied to   data collected from hundreds of people in controlled voting experiments, where people vote  after observing non-binding poll information.  
  We introduce a new  voting  model, the Attainability-Utility (AU) heuristic, which weighs the   popularity of  a  candidate according to the  poll,  with  the utility  of the candidate to the voter.  
  We argue that the AU model is cognitively plausible, and show that it is able to predict people's voting behavior 
  significantly better than other models from the literature. It was almost at par with (and sometimes better than) a machine learning algorithm  that uses substantially more information. 
   Our results provide new insights into the strategic considerations of voters, that undermine the prevalent assumptions of much  theoretical work in social choice. 
  \end{abstract}


\maketitle


 \section{Introduction}
Voting is a commonplace   tool for group decision making, used  in political elections,   in professional committees, in local assemblies, and  also  in online platforms such as \url{Doodle.com} and 
\url{robovote.org}. 
While there is general consensus that people vote strategically, understanding  individual 
voting behavior is a challenging open question. Due to inherent uncertainty about other people's votes, the strategies that people apply are far from obvious.

Researchers in economics, political science,  and more recently in AI and computational social choice,  have suggested various  models to represent and reason about   voters'  decision making 
 under uncertainty. 
These include models of \emph{utility maximization, heuristic, and bounded rational} (see below).  In a recent paper, Meir, Lev and Rosenschein~\cite{MLR14} suggested different criteria  for evaluating   models of strategic voting, that included ``theoretic criteria'' (such as generality and discriminative power among actions), ``behavioral criteria'' (such as cognitive plausibility), and ``scientific criteria'' (such as alignment with empirical data). Theoretical analysis of voting models are abundant (see Sec.~\ref{sec:lit_models}), and in this work we focus on the latter two kinds of criteria. \rmr{we tie back the results to these criteria in the conclusion}

\newpar{Research Goal}
The goal of the paper is to study strategic choices of human voters, and in particular to test how their individual behavior fits  different types of models. We use real world data  from controlled experiments in which human voters either faced a non-binding poll or played a strategic game versus other people.   
\kg{the below is standard in AAAI and ML papers, if you want to make a point that you differ from ECON then great, I recommend to remove and save space}
We follow Wright and Leyton-Brown~\shortcite{wright2010beyond}, who separate collected data from strategic games into training and test data, and compared the predictive power of strategic decision-making models based on their predictive performance on the test data. 
  If a certain model predicts well the behavior of many voters, this is an important indication for the plausibility of this model.Prediction is a standard evaluation metric in behavioral economics~\cite{brandstatter2006priority,erev2017anomalies}.
   Thus it should be considered in addition to its theoretical  properties, cognitive limitations of the voters and so on.
    By further analyzing which models succeed and when they fail, we hope to better understand the considerations that guide people's strategic choices.


\subsection{Theoretical Models} 
We briefly describe common approaches for modeling strategic voting behavior in the theoretical literature. 

\newpar{Expected utility maximization} A rational voter maximizes her expected utility with respect to a  probability distribution  over the actions of the other voters. 
The distribution itself may be given exogenously (e.g., by a poll), or derived via equilibrium analysis from the uncertain preferences of the other voters.
Such models were developed mainly in the economics literature and are sometimes known as the ``calculus of voting''~\cite{riker1968theory,merrill1981strategic,MW93}.
A somewhat different model was suggested by Bowman et al.~\shortcite{bowman2014potential} for voting on multiple binary issues. The model explicitly estimates the ``attainability'' of each issue (the probability it gets a majority of the votes), and uses this estimate to calculate the expected utility of every possible combinatorial vote. 

\newpar{Heuristic decision-making} A voter uses some  function that maps  any given situation to an action. The voter is not assumed to be rational, and may not even have a cardinal utility measure or an explicit probabilistic representation of the different outcomes.  For example, a voter following the 
$k$-pragmatist  heuristics behaves as if only the $k$ leading candidates  are participating~\cite{RE12}. 

\newpar{Bounded rationality} A voter makes a rational strategic decision based on a  subjective, rather than accurate, belief.  These models present a mid-point between utility maximization and heuristics. One example of such a model is \emph{local dominance}~\cite{MLR14}, which assumes that each voter derives a set of possible outcomes based on a poll, and then selects a non-dominated action within these outcomes. Similar probability-free approaches were followed in \cite{OLPRR16,endriss2016strategic}. 

\medskip 
 All the models we work with assume that  individual voters may behave differently,  but each one follows a deterministic, consistent  voting strategy. They are unable to perfectly explain or predict votes that have a random component or where voter behavior changes over time. 
Therefore, if the models can still explain the data, it would mean that noise and learning only play a secondary role in people's voting decisions. We go back to this point when analyzing the results.

Other models, such as \emph{quantal-response equilibrium} (QRE)~\cite{mckelvey1995quantal,mckelvey2006theory}, \emph{sampling equilibrium}~\cite{osborne2003sampling} or \emph{trembling hand equilibrium}~\cite{obraztsova2016trembling} assume voters act stochastically. Evaluation of such models is much easier on the aggregate rather than individual level (see below), and therefore they are outside the scope of the current paper. \kg{but you can still compute the prediction of stochastic models on the behavior of individual voters. I think 
you should just remove this paragraph.}
\rmr{reviewers explicitly ask about stochastic models. No point in hiding them. There is a good reason no one used them for individual prediction.}

\subsection{Previous Empirical Work} 
While the literature is abundant with voting experiments, the vast majority  analyze voter behavior in political or organizational elections
~\cite{abramson1992sophisticated,felsenthal1993empirical,blais2000calculus,regenwetter2007sophisticated,van2013vote}. These studies  test how well historic election results fit various game-theoretic models, without any consideration of individual votes. Further, each voter makes a single strategic decision, and her true preferences are typically unknown.

\newpar{Explaining aggregate voting behavior}
Some controlled experiments track voters' decisions in different situations (e.g., \cite{Forsythe1996ThreeCandidateExperiments,van2010strategic,tyszler2016information}).
 Most of these experiments included groups of 12-70 subjects who played a repeated strategic voting game, knowing the preferences of others but not how they are going to vote. Yet these papers focused on how well \emph{aggregate behavior} fits the equilibrium models. 
 For example, in the QRE model used in \citet{tyszler2016information}, voters are assumed to vote for low utility candidates with some probability, which is determined by a parameter of the model.
 A model is considered to be an adequate explanation for a dataset if there are some parameters that result in a similar distribution of votes (e.g., a similar rate of strategic compromise) to the one observed in the data. However, such models are not designed to track individual behavior and whether it is consistent. For example, a 20\% rate of strategic compromise could result either from a small group that is consistently strategic, from all voters being occasionally strategic, or even from some  random component in the behavior.


\newpar{Explaining individual voting behavior}
  Blaise et al.   compare individual behavior to rational models, inferring voters' parameters from verbal surveys~\cite{blais2000calculus} or from carefully designing the conditions of a controlled experiment~\cite{blais2014vote}. However, they focused on voting with two candidates, where the only strategic decision is whether to vote or abstain.

 We emphasize that all of the above work tested how well empirical data fits the theory \emph{in retrospect}, without dividing the data into separate training and test sets. This approach  may cause overfitting, especially in complex models with many parameters.


  Tal et al.~\shortcite{TalMG15}  study voter behavior under poll information,  
   but did not compare to any existing decision model, neither  suggested a new one. 
   They demonstrated empirically that there are voters exhibiting different behaviors: in particular,  truthful voters, voters who compromise strategically, and voters that tend to vote for the poll leader (``leader biased''). These findings were part of our inspiration to focus on understanding individual votes.
 

 \label{sec:contributions}
  Our contributions and results are as follows: 
 \begin{enumerate}
  \item We provide a new voting model called the  Attainability-Utility (AU) heuristic. 
 The  model  
 is based on the  model suggested by  Bowman et al.~\shortcite{bowman2014potential} which considers the attainability of a binary issue (the probability it is accepted in referendum) when computing the expected benefit of a vote. 
    The AU heuristic extends this model in two ways, first by considering multi-candidate voting settings, second by including a parameter that measures the tradeoff between 
  how much the voter values candidates' attainability given the poll information versus their \emph{utility} (if selected). 
 \item We collect the strategic decisions of 520  people in voting experiments with three candidates, where participants each play up to 36 rounds, each round with different poll information and preferences (more than 14,000 decisions in total). All of the data and code will become available for the public using repositories 
 such as \url{votelib.org}.

  
 \item  Using behavioral data from our experiments as well as from \citet{TalMG15} and \citet{tyszler2016information}, 
 we compare the performance of the AU approach to that of theoretical decision models from the literature and to benchmarks set by off-the-shelf machine learning algorithms. 

 
 
\item Our results show that the AU model outperforms all other voting models, some of them by a large margin, and gets close to the benchmark set by machine learning algorithms. In particular, AU is able to capture much of the behaviors described by the models in Sec.~\ref{sec:lit_models}. 
 Most errors in the prediction of the AU model can be attributed to participants who played few number of rounds, demonstrated random behavior, and/or changed their strategy during the experiment. 
\item Our main insight from the success of the AU model is that people  \emph{independently} evaluate each candidate and use simple substitutes to probabilistic calculations. These findings are in line with the more general research on decision making under uncertainty, and at odds with the underlying assumptions of most models from the social choice literature.  
\kg{the paragraph below talks about understanding, but doesn't say what we learned. 
 need to mention the models we used to explain people's behavior, e.g. leader bias, k pragmatics, LD,
 and that people reported using that behavior. 
when did people use one model, when did they use another? what were the 'types'? otherwise all we have is an empty declaration that we understand behavior}  

 \end{enumerate}
This is the first paper to provide an empirical evaluation  of theoretical decision making models on individual voter behavior under poll information, and the first to test the predictive performance of any voting model in general. Understanding the strategic decisions made by voters of different types, 
is crucial to  the development, analysis and application of  voting rules in strategic environments, and can inform the 
design of agents for making voting decisions with other people~\cite{yosef2017haste,Bitan2013SocialRankingsInHumanComputerCommittees}.

\section{Preliminaries}

\def\calU{\mathcal {U}}
\def\calD{\mathcal {D}}

We consider a single voter who faces a decision, to vote for one of several candidates $C$.  
 We use  the Plurality rule   which collects the total number of votes for each candidate, and returns the candidate(s) with the largest number of votes. 

The voter has a cardinal utility function $u:C\rightarrow \mathbb R$, where $u(c)$ is the utility of the voter if candidate $c$ wins (different utility for each candidate). 
In case of a tie with multiple winners $W\subseteq C$, the utility to the voter is 
 $u(W)=\frac{1}{|W|}\sum_{c\in W}u(c)$. Denote by $\calU(C)$ the set of all utility functions over the set $C$.

Prior to her vote, the voter is faced with non-binding poll information that reflects the popularity of each candidate. Formally, the poll is a vector $\vec s\in \mathbb N^m$, where $s(c)$ is the number of voters expected to vote for  $c$.  Denote  $n=\sum_{c\in C}s(c)$.

We index the candidates $q_1,q_2,\ldots$ from the perspective of the voter, where $q_1$ is the most preferred, then $q_2$, and so on.

\def\eps{\varepsilon}
\medskip
A \emph{decision model} (for Plurality with $m$ candidates and a poll) is a function $M:  \calU(C) \times \mathbb N^m \rightarrow C$. Here, $M(u,\vec s)\in C$ is the vote of a voter with utility function $u$, using decision model $M$ given a poll $\vec s$.  We use a superscript for the name of the decision model, and subscripts to denote voter-specific parameters, if relevant. 
 For example, a voter who is always truthful regardless of the poll follows the decision model $M^\truth(u, \vec s) := \argmax_{c\in C}u(c)$, which is $q_1$. 
 
\begin{figure} 
\centering
\begin{minipage}{0.61\columnwidth}
        \begin{figure}[H] \includegraphics[width=5.2cm]{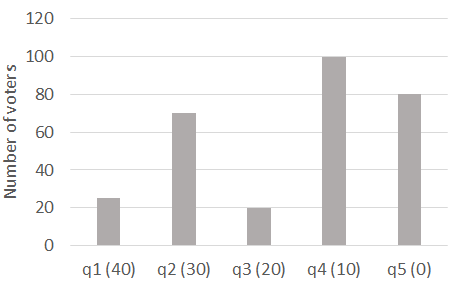} 
      \end{figure}
         \end{minipage}
     \begin{minipage}{0.32\columnwidth}
     \begin{small}
\begin{tabular}{l|c}
 Decision model   & vote\\
      \hline
      $\KP,  k=2$& $q_4$\\
      $\KP,  k=4$& $q_1$\\
      $\CV,   \eta=8$& $q_2$\\
      $\CV,  \eta=10000$& $q_4$\\
       $\LD,\  r=0.01$& $q_1$\\
      $\LD, \quad r=0.08$& $q_2$\\
      $\LDLB$, $r=0.01$& $q_4$\\
      $\LDLB$, $r=0.08$& $q_2$\\
    \end{tabular}
    \end{small}
    \end{minipage}
    \caption{\label{fig:pollExample}
 Left:   Poll $\vec s$ from Example~\ref{ex:poll}. The utility of each candidate to the voter appears in brackets, and the height of the column is the number of votes. Right:  The candidate selected by each decision model. \fsp}
\end{figure}

To illustrate we introduce a running example with 5 candidates, and specify which candidate the voter will choose under every decision model. 
\begin{example}
\label{ex:poll}
 The set of candidates is $C=\{q_1,\ldots,q_5\}$, 
  and the voter's utility is described by the vector $u=(40,30,20,10,0)$ (preferences are lexicographic). 
  Poll scores are given by $\vec s=(25,70,20,100,80)$, where $n=295$ voters. Figure~\ref{fig:pollExample} (left) shows the poll scores of all candidates graphically. 
\end{example} 


\subsection{Decision models from the literature}\label{sec:lit_models}

 We   describe  decision-making models of voting behavior  from the literature. 
 Figure~\ref{fig:pollExample} (right) 
 shows the result of using each of the decision models, applied to the voting decision in Example~\ref{ex:poll}.

\newparc{k-pragmatist (KP)}  Let $B_k(\vec s)$ contain the $k$ candidates with highest score in $\vec s$, Reijngoud and Endriss~\shortcite{RE12} formalized the $k$-pragmatist  heuristic (following early work such as \cite{brams1978approval} which  selects the most preferred candidate among $k$ candidates with highest   score in $B_k(\vec s)$:
$$M^\KP_k(u,\vec s) := \argmax_{c\in B_k(\vec s)}u(c).$$
We allow $k$ to be an individual parameter that differs from voter to voter. When  $k=1$  the rule always selects the leader of the poll, and for $k=m$, $M^\KP_m\equiv M^{Truth}$. 
In Figure~\ref{fig:pollExample}  for $k=2$ the voter will vote for  the candidate that is most  preferred among   the two leading candidates ($q_4$ and $q_5$).  For $k=4$, the voter considers all  candidates except $q_3$ as possible winners, and  will vote for her most preferred candidate $q_1$.


\newparc{Calculus of Voting (CV)}
The calculus of voting suggests that a rational voter  always votes in a way that maximizes her expected utility~\cite{riker1968theory,MW93}. The complications of the model usually arise from the fact that the voter is assumed to know  the other voters' {preferences}, and uses an equilibrium model to predict their votes. We consider a simpler version where the distribution of votes is given exogenously~\cite{merrill1981strategic}, as is the case with poll information.

We denote by  $\calD(\vec s)$ the  distribution on the actual candidate scores, conditional on poll scores  $\vec s$. 
We say  a voter is \emph{pivotal for candidate $y$ over $x$}, if voting for $y$ 
makes $y$ a joint or unique winner, whereas any other vote results in the victory of candidate $x$. Denote by  $P_{\vec s,\calD}(x,y)$ the probability that the voter is pivotal for $y$ 
over $x$  given the  distribution over 
candidate scores  $\calD$ induced by poll  $\vec s$.
A voter following the calculus of voting (CV) model maximizes her expected utility: 
$$M^{\CV}_\calD(u,\vec s) := \argmax_{c\in C}\sum_{c'\neq c}P_{\vec s,\calD}(c',c)(u(c)-u(c')).$$

To make the  model concrete, we  determine a specific distribution $\calD$  in a way that depends on the score of the candidates in the  poll $\vec{s}$. We use $P_{\vec s,\eta}$ as a shorthand for $P_{\vec s,\calD}$ when $\calD(\vec s)$ is a multinomial distribution with $\eta$ voters, and the probability for sampling a vote for each candidate $c$ is $s(c)/n$. 
When $\eta=n$ (i.e., the true number of voters), this means that $M^{\CV}_\eta$ selects the candidate that exactly maximizes the voter's expected utility 
 given the true distribution over candidate scores.
 However, the $M^\CV_\eta$ decision model allows for a more flexible, bounded-rational decision: when $\eta<n$ the voter overestimates her  true pivot probability, and thus her influence on the outcome, whereas $\eta>n$ means that she underestimates her influence.  In Figure~\ref{fig:pollExample} when $\eta=10000$ the resulting vote is $q_4$ and 
 when $\eta=8$ the resulting vote is $q_2$.


\newparc{Local Dominance  (LD)}
Under the Local dominance model~\cite{MLR14,Meir15}, a bounded-rational voter has an `uncertainty parameter' $r$. 
Meir et al.~\shortcite{MLR14} characterize the set of undominated candidates  $U(\vec s,u,r)$ in poll $\vec s$ for a voter with utility $u$ and parameter $r$:
 \begin{itemize}
 \item The set of Possible Winners $PW$ includes all candidates whose score in $\vec s$ is at least $\max_{c\in C}s(c)-2r\cdot n$.
 \item If $|PW|\geq 2$, then the undominated candidates are all candidates in $PW$ except the least preferred.
 \item If $|PW|=1$, then all candidates are undominated. 
 \end{itemize}

The decision model of such a voter selects the most preferred undominated candidate, if more than one exists:
$$M^{\LD}_r(u,\vec s) := \argmax_{c\in U(\vec s,u,r)}u(c).$$

In Figure~\ref{fig:pollExample} we see that for $r=0.01$ the voter believes that the poll is very accurate (the score of each candidate may change by at most $r\cdot n<3$ votes), and there is only one possible winner ($PW=\{q_4\}$). In this case, all candidates are undominated and the voter remains truthful ($M^\LD_{0.01}(u,\vec s) = q_1$). 
When $r=0.08$, the voter believes that the poll is not very accurate and  $PW=\{q_2,q_4,q_5\}$. In such a case both $q_2,q_4$ are undominated and $M^\LD_{0.08}(u,\vec s) = q_2$.

\newparc{Local-Dominance with Leader bias (LDLB)}
Inspired by the findings of Tal et al. \shortcite{TalMG15} on ``leader bias'',
 we modify the local dominance model to allow such behavior: when the voter is certain that there is only one possible winner ($|PW| = 1$), she simply votes for the leader (instead of truthfully), i.e., $M^{\LDLB}_r(u,\vec s) := M^{\LD}_r(u,\vec s)$, 
and otherwise $M^{\LDLB}_r(u,\vec s) :=PW$.
 In Figure~\ref{fig:pollExample} we see that this model acts similarly to the LD model. However when there is only one possible winner, a voter following the LDLB model will vote for the leader ($d$ in this case).  

\newparc{Attainability (AT)} Bowman et al.~\shortcite{bowman2014potential} provide a model for  voting over multiple binary issues. The  attainability   of issue $j$ is  a measure of  certainty  that the eventual number of votes cast for $j$ will reach the majority threshold required for approval. It is defined as 
$$\hat A_\beta(j,\vec s) := \frac{1}{\pi}\arctan(\beta\cdot(s_j-\frac12))+\frac12,$$
where $s_j$ is the expected number of votes in favor of issue $j$, and $\beta$ is a voter-specific parameter.


The ``candidates'' considered in Bowman et al.\shortcite{bowman2014potential} are all possible subsets of issues, i.e.,   $C=2^{\{1,\ldots,k\}}$, where w.l.o.g. the voter gains some nonnegative utility $u_j$ from each issue $j$ being approved. Then the utility of a candidate $c \subseteq \{1,\ldots,k\}$ is the sum of utilities of all issues in $c$, and its attainability  $\hat A_\beta(c,\vec s)$ is defined as the product of $\hat A_\beta(j,\vec s)$ for all $j\in c$, and $1-\hat A_\beta(j,\vec s)$ for all $j\notin c$. 
%
The voter 
 selects the candidate $c$ that maximizes the product of its attainability 
 and utility 
 ($\hat A_\beta(c,\vec s)\cdot u(c)$).

\begin{figure}[t]
\begin{center}
\includegraphics[scale=0.29]{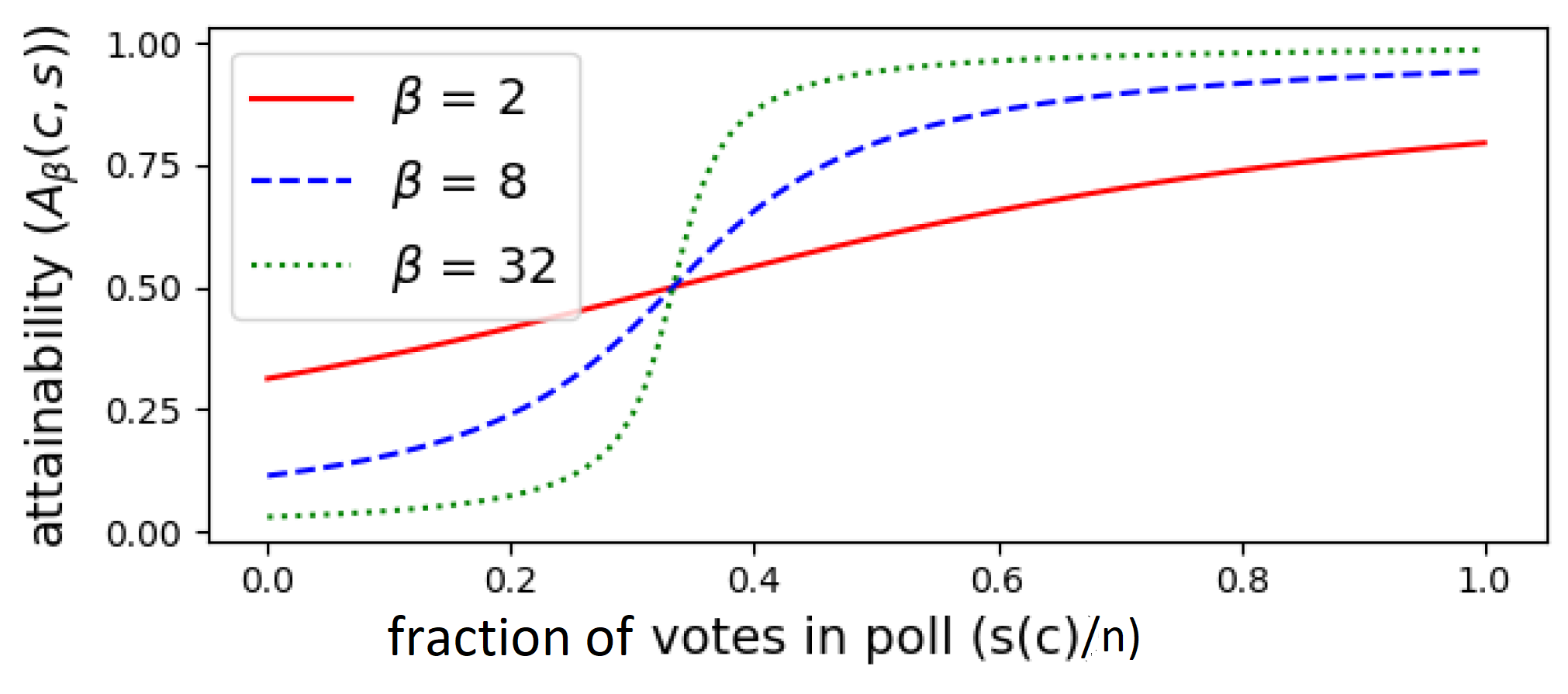}\fsp
\caption{\label{fig:attScore}The attainability  $A_\beta$ for different values of $\beta$ in a poll with 3 candidates. \vspace{-4mm}}
\end{center}
\end{figure}
 
 
To adapt the decision model to Plurality voting with $m$ candidates, we re-define the attainability function as $A_\beta(c,\vec s) := \frac{1}{\pi}\arctan(\beta\cdot(s(c)-\frac1m))+\frac12,$
and define the attainability choice function (AT) as 
$$M^{\AT}_\beta(u,\vec s) := \argmax_{c\in C}A_\beta(c,\vec s)\cdot u(c).$$

Figure~\ref{fig:attScore} shows how $\beta$ affects  the attainability score. Candidates that are tied have the same attainability. As shown by the figure, high $\beta$ means that a small advantage in score  translates to a large gap in attainability.

\section{The Attainability-Utility  (AU) model}
\label{sec:auv}
Bowman et al.'s AT model   allows voters some flexibility in how they estimate attainability using the $\beta$ parameter. 
However it  assumes the same model for each voter. 
We extend the attainability model by an  additional parameter that lets each voter choose a different tradeoff between the  attainability and utility of 
candidates. 
  To this end we define the \emph{Attainability-Utility} (AU) decision rule as 
$$M^{AU}_{\alpha,\beta}:=\argmax_{c\in C}\left((\eps+u(c))^\alpha \cdot (A_\beta(c,\vec s))^{2-\alpha}\right),$$
where $\eps$ is a small constant added to handle 0 utility ($\eps$ can also be used as a parameter to control the utility range).

\begin{figure}[t]
\fsp
\includegraphics[scale=0.5]{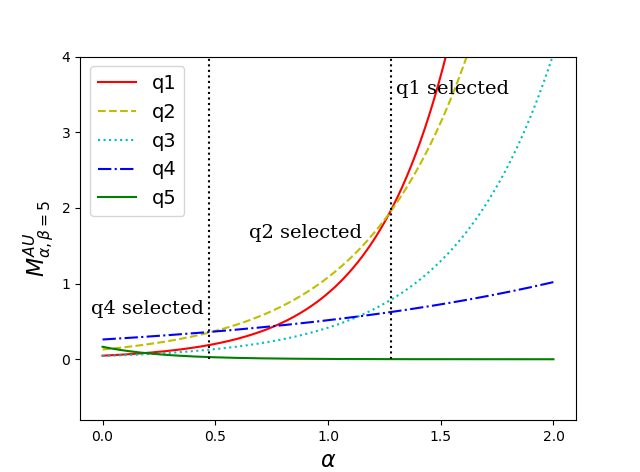}
\caption{ \label{fig:pollScore} 
The AU scores of all five candidates from Example~\ref{ex:poll} for $\beta=5$ and  different values of $\alpha$.\fsp}
\end{figure}

Intuitively, the $\alpha$ parameter trades-off the relative importance of attainability and utility, where $\alpha=0$ means the voter always selects the candidate with maximal score,  and $\alpha=2$ means the voter is always truthful. Figure~\ref{fig:pollScore} shows how the relative score (and the selected candidate) changes as we increase $\alpha$. When $\alpha$ is small, $\AVU$ will prefer $q_4$ as it has more votes (higher attainability) and when it is large the $\AVU$ will prefer $q_1$ as it got higher utility.  Note that we get the AT model as a special case when setting $\alpha=1, \eps=0$. We further discuss the meaning of these parameters in Sec.~\ref{sec:AU_cog}.


\kg{there was a discussion here on ties that seemed like a discussion between Reshef and Roy  had turned into paper text! I removed it}
 
\section{Methodology}
\label{sec:methodology}
\paragraph{Datasets}

 We evaluated the different models described above  on several datasets as follows.
 
 \begin{center}
 \resizebox{0.45\textwidth}{!}{
    \begin{tabular}{|l|cccc|}
    \hline
           & D32  & D36  & TMG15  & TS16 \\
           \hline
 \# participants  & 187  & 335  & 437     & 144 \\ 
 \# voters in poll& 1000 & 1000 & 8 to 10000 & 12 \\
\# rounds & up to 32 & up to 36 & up to 20 & 40 \\
 \#  instances  & 4886 & 9478 &  8011 &  5760 \\
 \hline
 
\end{tabular} 
 }
 \end{center}
  
\begin{figure}
\centering
\includegraphics[width=7.5cm]{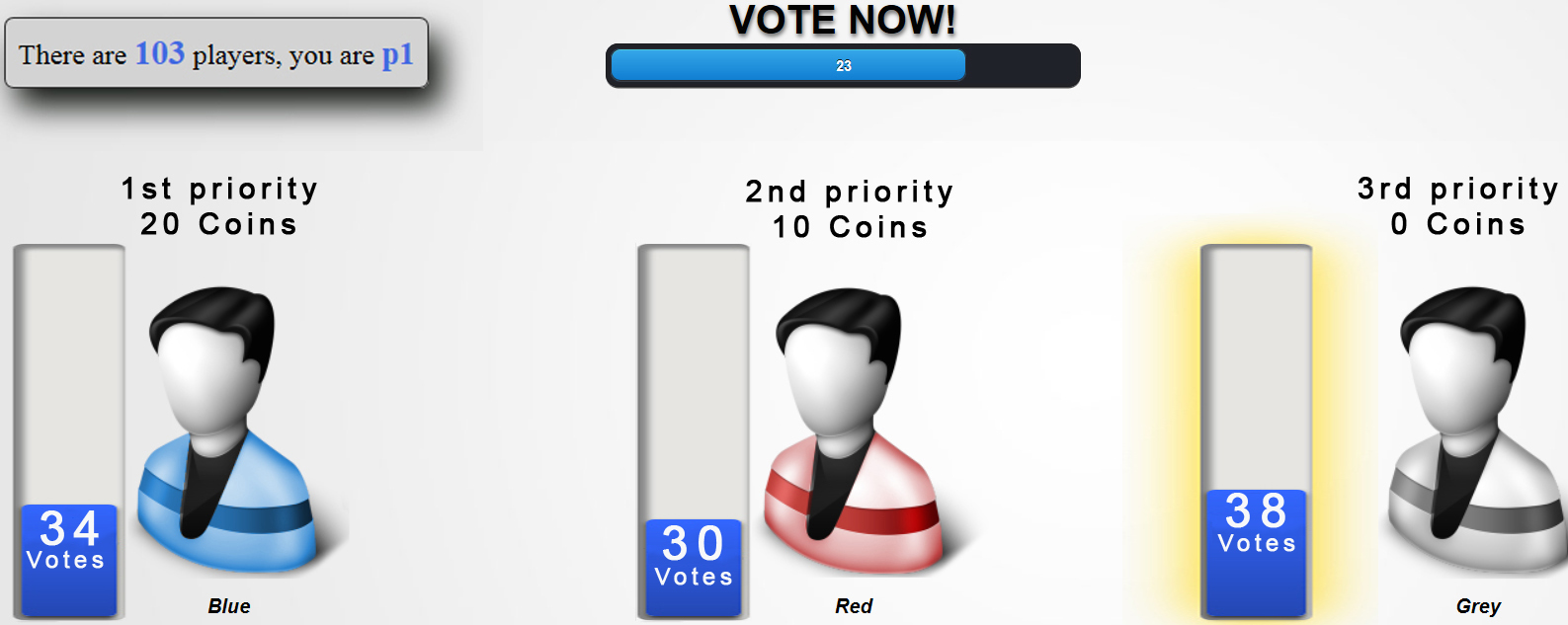}
\caption{A snapshot from the experiment (taken from \citet{TalMG15}). Blue is the most preferred candidate $q_1$, thus if Blue wins, the participant gets 20 coins ($10$ cents). The bars show the poll scores $s_1,s_2,s_3$. 
}
\label{fig:gameOne}
\end{figure}

 Three of the datasets (D32, D36 and TMG15) were collected 
 using the framework of Tal et al. \shortcite{TalMG15}, in which   voters played multiple one-shot voting rounds.  A snapshot of the GUI used for this setting is shown 
 in Figure~\ref{fig:gameOne}.
 
 Each round included a single human participant, that is automatically assigned preferences over candidates, observes a noisy ``poll'' with the expected votes of the entire population (e.g.,  1000 voters), and then votes once. 
 
 The outcome of the round was generated by sampling each of the other votes i.i.d using the poll scores as the distribution (e.g. in Fig.~\ref{fig:gameOne} we sample 102 ``voters", each of which votes Blue w.p. $\frac{34}{102}$). Participants were only informed that the poll was inaccurate, but not on the exact distribution. The final score of each candidate  and the outcome were shown in the end of each round. In all 
 datasets the reward for participants was determined by the 
 position of the winning candidate in their preferences, using the average reward if there was more than one winning candidate.

 We used this framework to generate datasets D32 and D36, presenting the participants with a different poll each time.  All participants were recruited via the Amazon Mechanical Turk platform.
 The reward was $R_j$ for each round where $q_j$ was elected, where $R_1>R_2>R_3$. For most of the participants, we set $R_1=10\cent, R_2=5\cent, R_3=0\cent$. For some participants we varied the rewards.
 
  The   
 dataset TS16 was generated by Tyzsler and Schram  \shortcite{tyszler2016information}. Here, every voting round was a 12-player complete information game with dictated preferences over 3 candidates, and the outcome was the result of all actual votes rather than artificial samples. 
 We used the other 11 voters' true top preferences (which are visible) as a true ``poll" input to the different decision models. 

 \medskip
 In all datasets, only when $q_1$ is ranked last at the poll, the voter may have a monetary incentive to  vote for $q_2$. There is never  a monetary incentive to vote for $q_3$.

\newpar{Random Forest (RF)  Benchmark}
We applied  off-the-shelf machine learning algorithm to build predictive models of voting behavior. 
We used two types of features: those relating to the particular voting round (examples: the gap $s_1-s_2$ between the two leaders of the poll, the number of votes $s_2$ in the poll, 
the winning candidate in the poll);
and those aggregating the behavior of the voter (examples: the frequency that the voter chose $q_1$, $q_2$ and $q_3$ in the training set,  the frequency of a strategic compromise, and the number of dominated actions). 
Using  these features, we compared the performance  of black-box prediction models  on  the D32 dataset.  

We compared  Random Forest, Neural Network, AdaBoost algorithms,   CART (Decision Tree), Support Vector Machines  and Logistics Regression.\footnote{We used the \texttt{sklearn} ensemble python package for this purpose~\cite{sklearn_api}.   The full list of features is available at \url{https://github.com/AdamLauz/OneShotVoting/blob/master/Documentation/One_Shot_ML_features_description.pdf}.} The best  performance was exhibited by a random forest ensemble model using 100 weak trees as subclassifiers, and a Gini splitting criterion. We thus used this algorithm (henceforth, \emph{RF}) as our benchmark.

 \newpar{Evaluation}
  We used a  ten-fold cross validation method. We divided the data of each voter into 10 folds (when possible).
 KP has only three parameter values. For the other models, we discretized the parameter space.
 
For each of the decision models KP, CV, AT, LD, LDLB and AU, 
we used a \emph{basic fitting procedure} to train each model 
separately for each voter: 9 folds of data for this voter were used to fit the parameters of the model, and applied the obtained model on the tenth fold to predict the voter's actions. 
 Since each voter has only few samples and the parameter space of each decision model is small, we used a brute-force search to find the best parameters for each model. 
 For example, for the LD model, we found for each voter the parameter $r$ such that $M_r^{LD}$ agrees with the largest number of rounds in the training set.
 
 The \emph{prediction error} of a model is the number of wrong predictions on the test set, divided by the total number of rounds.


\section{Results and Analysis}

\begin{figure}[t]
\begin{center}
\includegraphics[scale=0.42]{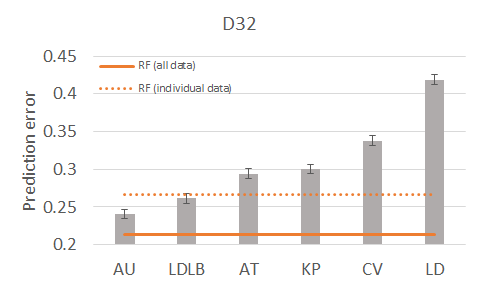}\!\!\!\!
\includegraphics[scale=0.42]{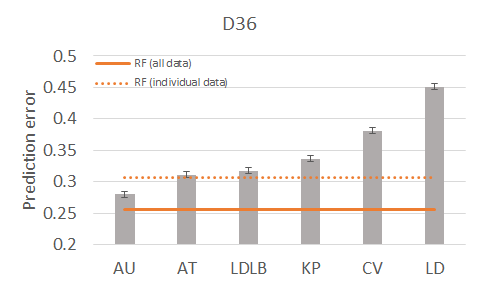}
\\
\includegraphics[scale=0.42]{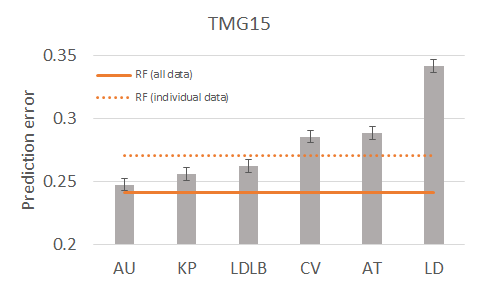}\!\!\!\!
\includegraphics[scale=0.42]{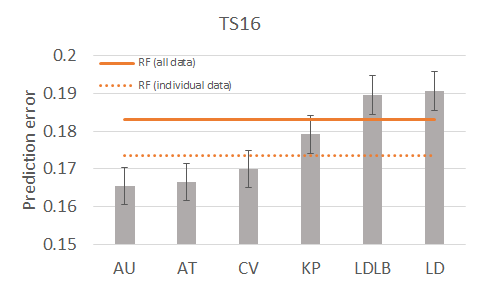}
\caption{\label{fig:f_all} Prediction error for each of the decision models on all four datasets. The horizontal orange lines mark the performance of the RF benchmark when trained on the entire data (solid) or restricted to individual data (dotted).\fsp}
\end{center}
\end{figure}
 
Figure~\ref{fig:f_all} shows the performance of all decision models on the datasets.  
We report the prediction error of the models, adding error bars of two standard deviations.

We  can see that the AU model outperforms all other decision models, with the LDLB model second, and the models that ignore leader bias (CV, LD) far behind.  These results are statistically significant in all datasets  ($p<0.05$) except in TS16, where there was no significant difference between the performance of AU and AT.   




\subsection{AU performance vs. the benchmark}
 Random Forest (RF) uses many features and can create an arbitrarily complicated model, which learns from the entire population of hundreds of voters rather than just from several individual samples. In addition, it uses temporal features and can thus in principle predict even behavior that changes over time.  This is why we use RF as a benchmark that is supposed to be hard to beat.
 Even so, RF does not perform uniformly better than the behavioral models.
 
\begin{figure}[t]
\begin{center}
\includegraphics[scale=0.45]{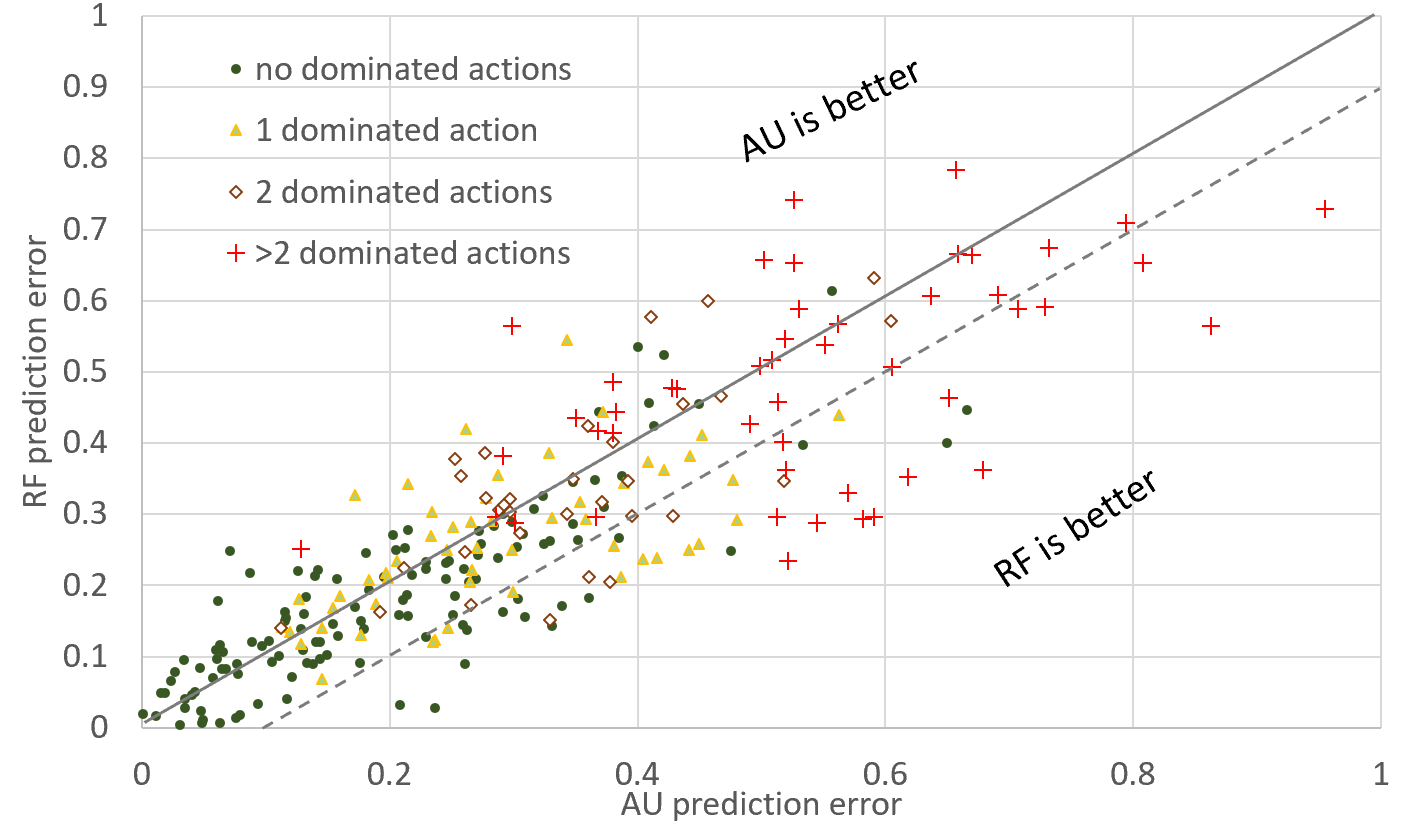}
\caption{\label{fig:scatter} Prediction error of AU versus RF for each voter. Only voters with at least 16 rounds are shown. The prediction of RF for voters below the dashed line was better by at least 10 percentage point.\fsp}
\end{center}
\end{figure}

 Figure~\ref{fig:scatter} breaks down 
the error of the AU model by individual voters (we refer to the different colors in the next subsection). \kg{important - if this figure remains then need to explain the colors} \rmr{we do in the next subsection}
It demonstrates visually that 
 AU beats RF for many individual voters (about a 100 out of 335), and that the advantage of RF is mainly due to a group of voters for which AU seems to perform substantially worse (those below the dashed line)\rf{This sentence doesn't sound correct. It says that RF succeed only for small number of voters, however its better for voters below the regular line, not the dashed line (which look like at least half). Under the dashed are voters that RF succeed marginally.}. Indeed it is possible that the AU model is appropriate for most voters but not for all (see  discussion in Sec.~\ref{sec:types}).

 Another factor is the data used for learning: while AU and all other decision models fit their parameters for a particular voter solely based on her own behavior in other rounds, the black-box algorithms had access to votes of other voters as well. When restricted to learn only from the samples that belong to the same individual, the error of RF leaped dramatically (see dotted lines in Fig.~\ref{fig:f_all}). This indicates that information about the entire population could be exploited to further improve the behavioral models.   In addition, the performance of RF (as well as the other black-box algorithms we tried) reduces more rapidly when we learn from a small or non-representative sample. 
 
\subsection{Where are the errors?}\label{sec:errors}
We analyze the factors that contributed most to prediction error, with a focus on  the AU model.

\newpar{Some voters  are harder to predict}
We say that a candidate is \emph{dominated} (in a particular round) if there is another candidate that is associated with a higher score in the poll and is also more preferred by the voter. E.g. $q_3$ in Example~\ref{ex:poll} is dominated by $q_2$. We count the number of dominated actions each voter performed throughout the experiment.  
Note that a dominated action is  never predicted by any of the decision models we considered. 
It is   hard to think of any rational justification for voting to a dominated candidate. We thus conjecture that dominated actions are indication for some random component in the behavior of the voter. 

We classified voters by the number of times they used a dominated action. In Fig.~\ref{fig:scatter} we can see that the number of dominated actions substantially affects prediction accuracy not just for AU but also for the benchmark RF (and in fact for all models). The prediction error of AU for voters who completely avoid dominated actions is less than 18\%, and increases to almost 50\% for voters with more than 2 dominated actions, indicating that their behavior is almost completely unpredictable. We emphasize that every additional dominated action results in \emph{more than one} (about 1.6-3) predictions errors. This, together with the low performance of RF, corroborates our conjecture that dominated actions are merely an indication for noisy or random voting patterns.

\begin{figure}[t]
\begin{center}
\includegraphics[scale=0.75]{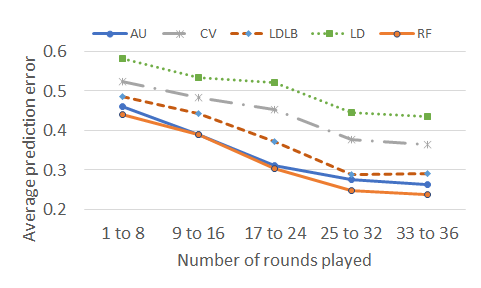}\fsp
\caption{\label{fig:roundsError}Prediction error as a function of the number of rounds in D36.  \fsp}
\end{center}
\end{figure}

Another factor that substantially affects prediction error is the number of rounds that a voter has played, where prediction error for voters who completed fewer rounds is much higher. See Fig.~\ref{fig:roundsError} which lists error as a function of number of rounds per voter. A likely explanation is that these voters are more prone to overfitting. 

\begin{figure}[t]
\begin{center}
\includegraphics[scale=0.8]{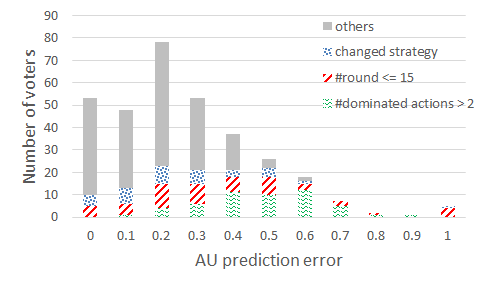}\fsp
\caption{\label{fig:VotersConsistencyFMeasure} A histogram of all D36 voters by their AU prediction error. We colored groups of voters for which certain conditions apply. 
\fsp}
\end{center}
\end{figure}

The histogram in Fig.~\ref{fig:VotersConsistencyFMeasure} shows that for almost all voters where AU had high error, the reason was random behavior (indicated by dominated actions), or few rounds. 


 
\medskip

\newpar{Behavior in some polls is hard to predict}
The behavior in polls that present the voter with an obvious dilemma (e.g., when her favorite candidate is trailing behind) is naturally harder to predict. In  Table~\ref{tab:by_poll}, we classified all polls into 6 poll types, based on the the order of candidates' popularity in the poll.

\begin{table}[ht!]
  \begin{center}
    \begin{tabular}{|cl|c|c|c|c|}
    \hline
     & poll type & D32 & D36 & TMG15 & TS16\\
      \hline
     & $q_1 > q_2 > q_3$ & 0.085 & 0.113 & 0.076 & 0.047\\
     & $q_1 > q_3 > q_2$ & 0.089 & 0.108 & 0.070 & 0.053\\
     & $q_2 > q_1 > q_3$ & 0.224 & 0.261 & 0.268 & 0.254\\
     & $q_3 > q_1 > q_2$ & 0.202 & 0.268 & 0.268 & 0.338\\
     & $q_2 > q_3 > q_2$ & 0.233 & 0.250 & 0.258 & 0.296\\
     $*$ & $q_3 > q_2 > q_1$ & 0.363 & 0.403 & 0.419 & 0.470\\
     \hline
    \end{tabular}
  \caption{\label{tab:by_poll}$\AVU$ error for each poll type. The order reflects the popularity of each candidate in the poll.\fsp}
  \end{center}

\end{table}

In the scenario most difficult to predict (where $s(q_3)>s(q_2)>s(q_1)$, marked with $*$), the poll order is reversed to the 
preference order of the candidates, and all three actions are frequently selected by the voters. For this case the prediction error of the AU model is above 35\% (and remains high even if we focus on voters who played all rounds and avoided dominated actions).  The results of the other models behaved similarly.  Note that while in Fig.~\ref{fig:f_all} prediction accuracy varies considerably between datasets, this is explained by the frequency of different poll types in each dataset (see Table~\ref{tab:by_poll}). 

We emphasize that   in TS16, most of the rounds people played as part of the majority group (see first two rows of Table~\ref{tab:by_poll}), and thus faced a trivial decision where all models predicted the same. 
This could be the reason  we did not obtain statistically significant results on this dataset, and also explains the poor performance of RF (which had few non-trivial rounds to learn from).

\begin{figure}[t]
\centering
\includegraphics[width=5.5cm]{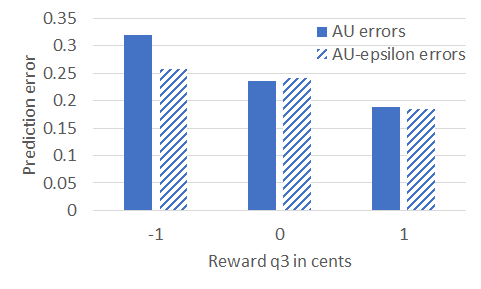}
\fsp
\caption{\label{fig:q3reward} AU and AU$_{\eps}$ accuracy in D32. AU$_{\eps}$ is identical to AU, except that $\eps$ is used as another parameter.\fsp } 
\end{figure}

\newpar{Negative reward is harder to predict}
In D32, we varied the reward $R_3$, to see the effect of positive/zero/negative reward (see Fig.~\ref{fig:q3reward}). 
Higher reward $R_3$ results in higher accuracy (the only statistically significant difference was between negative and zero reward, as we only varied the reward for 25 participants). 

\kg{can remove to save space, the eps model is not interesting enough}
A closer look revealed the reason for the excess failures: with a negative reward, $u_3+\eps$ is still negative, and thus the AU model would never select $q_3$. Perhaps surprisingly, participants do not care much about $R_3$ being negative. Adding $\eps$ as a third optimization parameter that may get higher values (so that $u_3+\eps>0$) completely negates that effect, as can be seen by the striped columns in Fig.~\ref{fig:q3reward}.  
We also varied the reward $R_2$ to get convex and concave utility scales rather than linear. While higher $R_2$ does lead to more frequent votes to $q_2$, we did not observe a consistent effect on prediction error. 

\subsection{Subjective reporting by participants}\label{sec:report}
From each participant in our experiment (datasets D32 and D36), we asked  to report their subjective answers  about how well they understood the instructions; which strategy they used in the study; and whether they changed a strategy during the game. 

\newpar{Descriptions of strategies} 
Some of the participants described strategies that are similar to the models we tested from the
voting literature. 
Some  primary examples appear below.
\begin{itemize}
\item  ``I tried to vote for the person most likely to beat the candidate that would give me no coins." Describes \KP{} with $k=2$. 
\item 
\emph{``I voted for either my first or second priority candidate.  I was more likely to vote for the one that appeared to have the highest probability of winning."}
- Describe behavior similar to \AT{}.
\item \emph{``My strategy was to mainly vote for who was leading except when it was a close race and then I voted for who would earn me the most points.''} - describes \LDLB{} with low $r$. 
\end{itemize}
Interestingly, people did not adhere to their reported strategies in all rounds, and often their behavior was predicted more accurately by a different model than the one they verbally describe.

For example, some people who explicitly declared that they would not vote for the least 
preferred candidate  did in fact choose this option. 
It is not clear whether this results from noisy behavior, from changing the behavior over time, or from poor self-reflection.


\newparn{Do voters use consistent strategies?}
Identifying individual changes in strategy from the data is very difficult with only a handful of samples per voter. However from the subjective self-reports, about 63\% of those who responded in D36, answered that they did not change their strategy, whereas only 18\% did.\footnote{The others provided an answer that could not be easily classified, e.g. ``It took me a few rounds to get the hang of it.''}

There was a strong correlation between how well people understood the instructions (by their self-report) and their consistency: in D36,  more than 80\% of those who reported perfect understanding, claimed they kept their strategy. Results in D32 were similar.

Recall the questions from the introduction about consistency and predictability.
The level of (reported) consistency strongly affected the empirical error: the average error of AU for consistent voters in D36 was about 24\% vs. 30\% for inconsistent ones (and 22\% vs. 34\% in D32). This can also be partly seen in Fig.~\ref{fig:VotersConsistencyFMeasure}, where voters who reported strategy change (dotted blue) are responsible for slightly more errors.  

\section{Discussion}
Regenwetter et al.~\shortcite{regenwetter2007sophisticated} observe that~~~~~
``\emph{...individual choice research finds actors to behave \textbf{worse than} normative theory requires, whereas the sparse empirical research on social choice appears to suggest that electorates may \textbf{outperform} normative expectations}.'' 
However,  most research they refer to considered \emph{aggregated behavior}, as discussed in the early sections.


Our Attainability-Utility (AU) model explains  well (and in particular much better than calculus of voting) the behavior of most subjects in the data, except those with inherent inconsistencies in their actions. This partly settles the discrepancy observed by Regenwetter et al.: on the individual level, most voters follow AU or other   heuristics that do not maximize expected utility, just like decision makers in other domains, even if on the aggregate level the vote distribution can be explained by more rational theories like calculus of voting~\cite{Forsythe1996ThreeCandidateExperiments,blais2000calculus}, or quantal response equilibrium~\cite{tyszler2016information}. Interestingly, quantal response can account for the frequency of dominated actions at the aggregate level, even if it cannot predict when a particular action will be dominated. We may therefore get a more complete picture of voters' behavior by combining individual and aggregate analysis (see also future work below).  

\subsection{Is AU cognitively plausible?}\label{sec:AU_cog}
\kg{great paragraph; need to add to the intro that we provide a justification 
for behavior from classical results in behavioral decision making and cognitive science}
There are two seemingly ``irrational'' components in the AU model (both inherited from Bowman~\cite{bowman2014potential}), that become apparent when we compare it to the ``rational'' Calculus of Voting method.
The first is the fact that the voter asses the chances of each \emph{candidate to win}, rather than of each possible \emph{tie}. The second is that this chance is estimated using a somewhat arbitrary transformation of the candidate's score (the logit-shaped ``attainability function''), rather than by explicit probabilistic calculations.

Both observations are much less surprising when we recall Kahnemann and Tversky's account of judgment under uncertainty~\cite{tversky1974judgment}: they explain that people often use simple substitutes for probabilistic calculation, that require low cognitive effort. For example, rely on how \emph{representative} each event is (in our case, the score of each candidate in the poll).\footnote{We tried a variation of the AU model, where attainability was replaced with the actual winning probability. This did not improve the performance of the model.}
Future experiments can further test this hypothesis by making candidates more prominent in other ways than higher score (e.g., using graphic features), and see if voters' behavior can still be explained when we translate this to greater attainability. The extensive literature on the various heuristics people use to evaluate likely outcomes (e.g.~\cite{bar1973subjective,chater2006probabilistic}) can also be used to develop better models of voting behavior.
\rmr{what we need to do next (not now) is to go back to Bar-Hillel paper and see if we can extract some better `attainability` function from her insights. She showed people polls and measured their similarity - we could use that.}

It is worth mentioning that trying to use a simple substitute for probabilistic calculations was the main motivation behind the Local Dominance model~\cite{MLR14}, but Local Dominance (like the Calculus of Voting) still focuses on ties. The fact that AU better explains the behavior of most voters (and in particular that LD fails to predict leader-biased actions) suggests that perhaps even LD is too cognitively prohibiting. 
Indeed, except for KP (which is perhaps too simple), AU is the cognitively easiest heuristic to apply, as it \emph{independently} evaluates each candidate. We note that the differences in cognitive burden become even more accentuated in elections with more candidates. We therefore expect the differences in performance to become more significant as well, and are currently collecting more data to test this hypothesis.

\subsection{Are there voters of different types?}\label{sec:types} 
\begin{figure}[t]
\begin{center}
\includegraphics[scale=0.4]{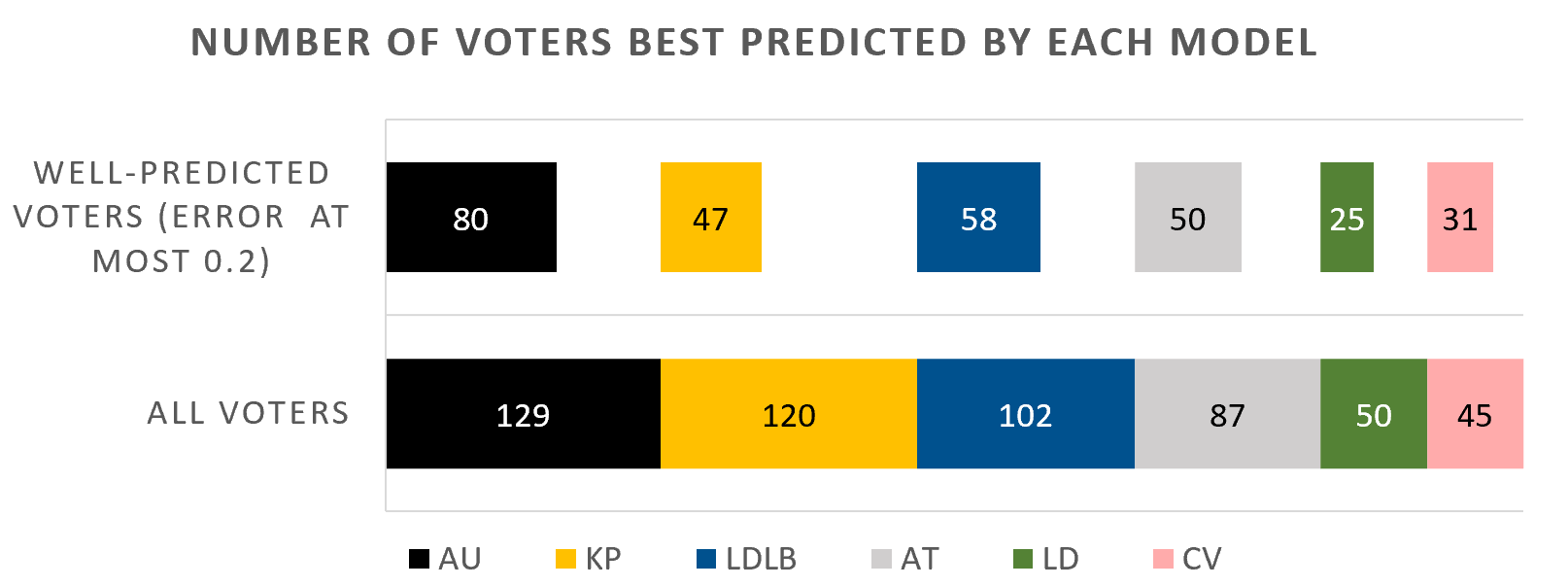}
\caption{\label{fig:best_model} 
The bottom bar shows, for each model, the number of participants in D32 and D36 for which this model achieved the best accuracy (possibly tied with other models).
The top bar shows the same information, for the subset of voters for which the best prediction error was at most $0.2$.\fsp}
\end{center}
\end{figure}

\kg{describe also in intro;}\rmr{I think this would hurt our message. I said we discuss it}While AU had the best performance overall,
there are many individual voters that are better predicted by one of the other models (not necessarily the leading one). 
This can be seen in the bottom bar in  Figure~\ref{fig:best_model} which shows the number of voters that were optimally predicted by each model.  In case of ties we `split' the voter among all leading models.
To illustrate, although 129 voters were best predicted by the AU heuristic, almost as many (120 voters) were best predicted by 
the KP heuristic.

At this point we face a dilemma when trying to explain the reason that so many voters are better predicted by other models:
One hypothesis (H1) is that AU can in principle account for the behavior of all voters, but is overfitting its parameters due to the small dataset of each voter. 
An alternative hypothesis (H2) is that there are indeed voters with different inherent behaviors that are better captured by other models, such as LDLB, KP and so on.

Some evidence for H2 is in the self reports where participants described distinct strategies. 
However, we believe there is stronger evidence for H1: first, the self reports are often inconsistent with the actual behavior, and AU in fact predicts well many of the voters who described specific strategies.
Also, AU can in principle explain (for some parameter values) almost all voters in the data, but we often fail to select the optimal parameters due to the small training sample so other models have fewer prediction errors. This can be seen in Fig.~\ref{fig:roundsError} where AU improves faster as voters have more samples to train on.
Lastly, the advantage of AU becomes more clear once we focus on voters with low error (top bar in Fig.~\ref{fig:best_model}). These are the voters for which the selected model is more meaningful.

To better answer this question, richer datasets that better distinct between decision models should be generated. 


\subsection{Discussion and future work}\label{sec:future}
Finding a model that perfectly explains the behavior of all voters is probably impossible.  
Yet, our AU model does well both on the ``behavioral'' and on the ``scientific'' criteria presented in \cite{MLR14}:  It is a fairly simple and cognitively plausible model, that captures the behavior of most voters well enough to predict their individual actions in various situations, and even to compete with machine-learning algorithms that use hundreds of features from the entire population. This model trades-off the popularity of a candidate (as a proxy for its winning chances) and its utility to the voter.  


Future  voting models should be extended to allow behavior that changes over time in some predictable way.
More importantly, deterministic decision models should be combined with stochastic ones like quantal response and trembling hand perfection~\cite{mckelvey2006theory,obraztsova2016trembling} to explain both consistent individual choices and random departures from those choices. New evaluation methods are needed for these combined aggregate and individual choices.

Our findings can inform the  development new and better models for strategic voting, much like the PrefLib project~\cite{prefLib2013} is contributing to the study of preference structure, as well as to inform the 
design of agents for making voting decisions with other people, which is a growing area of research~\cite{yosef2017haste,Bitan2013SocialRankingsInHumanComputerCommittees}.

Most of the decision models we used, including the new AU heuristic, naturally extend to more candidates and other voting rules. We  intend to run
experiments in more diverse settings (e.g. more than 3 candidates). Those experiments can expose  behaviors that do not exist in the  current data, can help differentiate between the decision models, and serve as a benchmark for the development of new models.
 
 
 \section{Acknowledgements}
 This work was supported in part thanks to the Israeli Science Foundation grant number 773/16. Thanks to Tyszler and Schram for making their data available to us for analysis.
 \bibliographystyle{ACM-Reference-Format}  
 \balance
\bibliography{main.bib}


\begin{thebibliography}{00}


\ifx \showCODEN    \undefined \def \showCODEN     #1{\unskip}     \fi
\ifx \showDOI      \undefined \def \showDOI       #1{#1}\fi
\ifx \showISBNx    \undefined \def \showISBNx     #1{\unskip}     \fi
\ifx \showISBNxiii \undefined \def \showISBNxiii  #1{\unskip}     \fi
\ifx \showISSN     \undefined \def \showISSN      #1{\unskip}     \fi
\ifx \showLCCN     \undefined \def \showLCCN      #1{\unskip}     \fi
\ifx \shownote     \undefined \def \shownote      #1{#1}          \fi
\ifx \showarticletitle \undefined \def \showarticletitle #1{#1}   \fi
\ifx \showURL      \undefined \def \showURL       {\relax}        \fi
\providecommand\bibfield[2]{#2}
\providecommand\bibinfo[2]{#2}
\providecommand\natexlab[1]{#1}
\providecommand\showeprint[2][]{arXiv:#2}

\bibitem[\protect\citeauthoryear{Abramson, Aldrich, Paolino, and
  Rohde}{Abramson et~al\mbox{.}}{1992}]%
        {abramson1992sophisticated}
\bibfield{author}{\bibinfo{person}{Paul~R Abramson}, \bibinfo{person}{John~H
  Aldrich}, \bibinfo{person}{Phil Paolino}, {and} \bibinfo{person}{David~W
  Rohde}.} \bibinfo{year}{1992}\natexlab{}.
\newblock \showarticletitle{"Sophisticated" voting in the 1988 presidential
  primaries}.
\newblock \bibinfo{journal}{{\em American Political Science Review\/}}
  \bibinfo{volume}{86}, \bibinfo{number}{1} (\bibinfo{year}{1992}),
  \bibinfo{pages}{55--69}.
\newblock


\bibitem[\protect\citeauthoryear{Bar-Hillel}{Bar-Hillel}{1973}]%
        {bar1973subjective}
\bibfield{author}{\bibinfo{person}{Maya Bar-Hillel}.}
  \bibinfo{year}{1973}\natexlab{}.
\newblock \showarticletitle{On the subjective probability of compound events}.
\newblock \bibinfo{journal}{{\em Organizational behavior and human
  performance\/}} \bibinfo{volume}{9}, \bibinfo{number}{3}
  (\bibinfo{year}{1973}), \bibinfo{pages}{396--406}.
\newblock


\bibitem[\protect\citeauthoryear{Bitan, Gal, Kraus, Dokow, and Azaria}{Bitan
  et~al\mbox{.}}{2013}]%
        {Bitan2013SocialRankingsInHumanComputerCommittees}
\bibfield{author}{\bibinfo{person}{Moshe Bitan}, \bibinfo{person}{Ya'akov Gal},
  \bibinfo{person}{Sarit Kraus}, \bibinfo{person}{Elad Dokow}, {and}
  \bibinfo{person}{Amos Azaria}.} \bibinfo{year}{2013}\natexlab{}.
\newblock \showarticletitle{Social Rankings in Human-Computer Committees}. In
  \bibinfo{booktitle}{{\em AAAI}}.
\newblock


\bibitem[\protect\citeauthoryear{Blais, Pilet, Van~der Straeten, Laslier, and
  H{\'e}roux-Legault}{Blais et~al\mbox{.}}{2014}]%
        {blais2014vote}
\bibfield{author}{\bibinfo{person}{Andr{\'e} Blais},
  \bibinfo{person}{Jean-Benoit Pilet}, \bibinfo{person}{Karine Van~der
  Straeten}, \bibinfo{person}{Jean-Fran{\c{c}}ois Laslier}, {and}
  \bibinfo{person}{Maxime H{\'e}roux-Legault}.}
  \bibinfo{year}{2014}\natexlab{}.
\newblock \showarticletitle{To vote or to abstain? An experimental test of
  rational calculus in first past the post and {PR} elections}.
\newblock \bibinfo{journal}{{\em Electoral studies\/}}  \bibinfo{volume}{36}
  (\bibinfo{year}{2014}), \bibinfo{pages}{39--50}.
\newblock


\bibitem[\protect\citeauthoryear{Blais, Young, and Lapp}{Blais
  et~al\mbox{.}}{2000}]%
        {blais2000calculus}
\bibfield{author}{\bibinfo{person}{Andr{\'e} Blais}, \bibinfo{person}{Robert
  Young}, {and} \bibinfo{person}{Miriam Lapp}.}
  \bibinfo{year}{2000}\natexlab{}.
\newblock \showarticletitle{The calculus of voting: An empirical test}.
\newblock \bibinfo{journal}{{\em European Journal of Political Research\/}}
  \bibinfo{volume}{37}, \bibinfo{number}{2} (\bibinfo{year}{2000}),
  \bibinfo{pages}{181--201}.
\newblock


\bibitem[\protect\citeauthoryear{Bowman, Hodge, and Yu}{Bowman
  et~al\mbox{.}}{2014}]%
        {bowman2014potential}
\bibfield{author}{\bibinfo{person}{Clark Bowman}, \bibinfo{person}{Jonathan~K
  Hodge}, {and} \bibinfo{person}{Ada Yu}.} \bibinfo{year}{2014}\natexlab{}.
\newblock \showarticletitle{The potential of iterative voting to solve the
  separability problem in referendum elections}.
\newblock \bibinfo{journal}{{\em Theory and decision\/}} \bibinfo{volume}{77},
  \bibinfo{number}{1} (\bibinfo{year}{2014}), \bibinfo{pages}{111--124}.
\newblock


\bibitem[\protect\citeauthoryear{Brams and Fishburn}{Brams and
  Fishburn}{1978}]%
        {brams1978approval}
\bibfield{author}{\bibinfo{person}{Steven~J Brams} {and}
  \bibinfo{person}{Peter~C Fishburn}.} \bibinfo{year}{1978}\natexlab{}.
\newblock \showarticletitle{Approval voting}.
\newblock \bibinfo{journal}{{\em American Political Science Review\/}}
  \bibinfo{volume}{72}, \bibinfo{number}{3} (\bibinfo{year}{1978}),
  \bibinfo{pages}{831--847}.
\newblock


\bibitem[\protect\citeauthoryear{Brandst{\"a}tter, Gigerenzer, and
  Hertwig}{Brandst{\"a}tter et~al\mbox{.}}{2006}]%
        {brandstatter2006priority}
\bibfield{author}{\bibinfo{person}{Eduard Brandst{\"a}tter},
  \bibinfo{person}{Gerd Gigerenzer}, {and} \bibinfo{person}{Ralph Hertwig}.}
  \bibinfo{year}{2006}\natexlab{}.
\newblock \showarticletitle{The Priority Heuristic: Making Choices Without
  Trade-Offs}.
\newblock \bibinfo{journal}{{\em Psychological Review\/}}
  \bibinfo{volume}{113}, \bibinfo{number}{2} (\bibinfo{year}{2006}),
  \bibinfo{pages}{409--432}.
\newblock


\bibitem[\protect\citeauthoryear{Buitinck, Louppe, Blondel, Pedregosa, Mueller,
  Grisel, Niculae, Prettenhofer, Gramfort, Grobler, Layton, VanderPlas, Joly,
  Holt, and Varoquaux}{Buitinck et~al\mbox{.}}{2013}]%
        {sklearn_api}
\bibfield{author}{\bibinfo{person}{Lars Buitinck}, \bibinfo{person}{Gilles
  Louppe}, \bibinfo{person}{Mathieu Blondel}, \bibinfo{person}{Fabian
  Pedregosa}, \bibinfo{person}{Andreas Mueller}, \bibinfo{person}{Olivier
  Grisel}, \bibinfo{person}{Vlad Niculae}, \bibinfo{person}{Peter
  Prettenhofer}, \bibinfo{person}{Alexandre Gramfort}, \bibinfo{person}{Jaques
  Grobler}, \bibinfo{person}{Robert Layton}, \bibinfo{person}{Jake VanderPlas},
  \bibinfo{person}{Arnaud Joly}, \bibinfo{person}{Brian Holt}, {and}
  \bibinfo{person}{Ga{\"{e}}l Varoquaux}.} \bibinfo{year}{2013}\natexlab{}.
\newblock \showarticletitle{{API} design for machine learning software:
  experiences from the scikit-learn project}. In \bibinfo{booktitle}{{\em ECML
  PKDD Workshop: Languages for Data Mining and Machine Learning}}.
  \bibinfo{pages}{108--122}.
\newblock


\bibitem[\protect\citeauthoryear{Chater, Tenenbaum, and Yuille}{Chater
  et~al\mbox{.}}{2006}]%
        {chater2006probabilistic}
\bibfield{author}{\bibinfo{person}{Nick Chater}, \bibinfo{person}{Joshua~B
  Tenenbaum}, {and} \bibinfo{person}{Alan Yuille}.}
  \bibinfo{year}{2006}\natexlab{}.
\newblock \showarticletitle{Probabilistic models of cognition: Conceptual
  foundations}.
\newblock \bibinfo{journal}{{\em Trends in Cognitive Sciences\/}}
  \bibinfo{volume}{10}, \bibinfo{number}{7} (\bibinfo{year}{2006}),
  \bibinfo{pages}{287--291}.
\newblock


\bibitem[\protect\citeauthoryear{Endriss, Obraztsova, Polukarov, and
  Rosenschein}{Endriss et~al\mbox{.}}{2016}]%
        {endriss2016strategic}
\bibfield{author}{\bibinfo{person}{Ulle Endriss}, \bibinfo{person}{Svetlana
  Obraztsova}, \bibinfo{person}{Maria Polukarov}, {and}
  \bibinfo{person}{Jeffrey~S Rosenschein}.} \bibinfo{year}{2016}\natexlab{}.
\newblock \showarticletitle{Strategic voting with incomplete information}. In
  \bibinfo{booktitle}{{\em Proceedings of the Twenty-Fifth International Joint
  Conference on Artificial Intelligence}}. AAAI Press,
  \bibinfo{pages}{236--242}.
\newblock


\bibitem[\protect\citeauthoryear{Erev, Ert, Plonsky, Cohen, and Cohen}{Erev
  et~al\mbox{.}}{2017}]%
        {erev2017anomalies}
\bibfield{author}{\bibinfo{person}{Ido Erev}, \bibinfo{person}{Eyal Ert},
  \bibinfo{person}{Ori Plonsky}, \bibinfo{person}{Doron Cohen}, {and}
  \bibinfo{person}{Oded Cohen}.} \bibinfo{year}{2017}\natexlab{}.
\newblock \showarticletitle{From anomalies to forecasts: Toward a descriptive
  model of decisions under risk, under ambiguity, and from experience.}
\newblock \bibinfo{journal}{{\em Psychological review\/}}
  \bibinfo{volume}{124}, \bibinfo{number}{4} (\bibinfo{year}{2017}),
  \bibinfo{pages}{369}.
\newblock


\bibitem[\protect\citeauthoryear{Felsenthal, Maoz, and Rapoport}{Felsenthal
  et~al\mbox{.}}{1993}]%
        {felsenthal1993empirical}
\bibfield{author}{\bibinfo{person}{Dan~S Felsenthal}, \bibinfo{person}{Zeev
  Maoz}, {and} \bibinfo{person}{Amnon Rapoport}.}
  \bibinfo{year}{1993}\natexlab{}.
\newblock \showarticletitle{An empirical evaluation of six voting procedures:
  do they really make any difference?}
\newblock \bibinfo{journal}{{\em British Journal of Political Science\/}}
  \bibinfo{volume}{23}, \bibinfo{number}{1} (\bibinfo{year}{1993}),
  \bibinfo{pages}{1--27}.
\newblock


\bibitem[\protect\citeauthoryear{Forsythe, Rietz, Myerson, and Weber}{Forsythe
  et~al\mbox{.}}{1996}]%
        {Forsythe1996ThreeCandidateExperiments}
\bibfield{author}{\bibinfo{person}{Robert Forsythe}, \bibinfo{person}{Thomas
  Rietz}, \bibinfo{person}{Roger Myerson}, {and} \bibinfo{person}{Robert
  Weber}.} \bibinfo{year}{1996}\natexlab{}.
\newblock \showarticletitle{An experimental study of voting rules and polls in
  three candidate elections}.
\newblock \bibinfo{journal}{{\em International Journal of Game Theory\/}}
  \bibinfo{volume}{25}, \bibinfo{number}{3} (\bibinfo{year}{1996}),
  \bibinfo{pages}{355--383}.
\newblock


\bibitem[\protect\citeauthoryear{Mattei and Walsh}{Mattei and Walsh}{2013}]%
        {prefLib2013}
\bibfield{author}{\bibinfo{person}{Nicholas Mattei} {and} \bibinfo{person}{Toby
  Walsh}.} \bibinfo{year}{2013}\natexlab{}.
\newblock \showarticletitle{PrefLib: A Library for Preferences
  http://www.preflib.org}.
\newblock In \bibinfo{booktitle}{{\em ADT'13}}. \bibinfo{pages}{259--270}.
\newblock


\bibitem[\protect\citeauthoryear{McKelvey and Palfrey}{McKelvey and
  Palfrey}{1995}]%
        {mckelvey1995quantal}
\bibfield{author}{\bibinfo{person}{Richard~D McKelvey} {and}
  \bibinfo{person}{Thomas~R Palfrey}.} \bibinfo{year}{1995}\natexlab{}.
\newblock \showarticletitle{Quantal response equilibria for normal form games}.
\newblock \bibinfo{journal}{{\em Games and economic behavior\/}}
  \bibinfo{volume}{10}, \bibinfo{number}{1} (\bibinfo{year}{1995}),
  \bibinfo{pages}{6--38}.
\newblock


\bibitem[\protect\citeauthoryear{McKelvey and Patty}{McKelvey and
  Patty}{2006}]%
        {mckelvey2006theory}
\bibfield{author}{\bibinfo{person}{Richard~D McKelvey} {and}
  \bibinfo{person}{John~W Patty}.} \bibinfo{year}{2006}\natexlab{}.
\newblock \showarticletitle{A theory of voting in large elections}.
\newblock \bibinfo{journal}{{\em Games and Economic Behavior\/}}
  \bibinfo{volume}{57}, \bibinfo{number}{1} (\bibinfo{year}{2006}),
  \bibinfo{pages}{155--180}.
\newblock


\bibitem[\protect\citeauthoryear{Meir}{Meir}{2015}]%
        {Meir15}
\bibfield{author}{\bibinfo{person}{Reshef Meir}.}
  \bibinfo{year}{2015}\natexlab{}.
\newblock \showarticletitle{Plurality Voting under Uncertainty}. In
  \bibinfo{booktitle}{{\em AAAI'15}}.
\newblock


\bibitem[\protect\citeauthoryear{Meir, Lev, and Rosenschein}{Meir
  et~al\mbox{.}}{2014}]%
        {MLR14}
\bibfield{author}{\bibinfo{person}{Reshef Meir}, \bibinfo{person}{Omer Lev},
  {and} \bibinfo{person}{Jeffrey~S. Rosenschein}.}
  \bibinfo{year}{2014}\natexlab{}.
\newblock \showarticletitle{A Local-dominance theory of voting equilibria}. In
  \bibinfo{booktitle}{{\em ACM-EC'14}}. \bibinfo{pages}{313--330}.
\newblock


\bibitem[\protect\citeauthoryear{Merrill}{Merrill}{1981}]%
        {merrill1981strategic}
\bibfield{author}{\bibinfo{person}{Samuel Merrill}.}
  \bibinfo{year}{1981}\natexlab{}.
\newblock \showarticletitle{Strategic decisions under one-stage multi-candidate
  voting systems}.
\newblock \bibinfo{journal}{{\em Public Choice\/}} \bibinfo{volume}{36},
  \bibinfo{number}{1} (\bibinfo{year}{1981}), \bibinfo{pages}{115--134}.
\newblock


\bibitem[\protect\citeauthoryear{Myerson and Weber}{Myerson and Weber}{1993}]%
        {MW93}
\bibfield{author}{\bibinfo{person}{Roger~B. Myerson} {and}
  \bibinfo{person}{Robert~J. Weber}.} \bibinfo{year}{1993}\natexlab{}.
\newblock \showarticletitle{A Theory of Voting Equilibria}.
\newblock \bibinfo{journal}{{\em The American Political Science Review\/}}
  \bibinfo{volume}{87}, \bibinfo{number}{1} (\bibinfo{year}{1993}),
  \bibinfo{pages}{102--114}.
\newblock


\bibitem[\protect\citeauthoryear{Obraztsova, Lev, Polukarov, Rabinovich, and
  Rosenschein}{Obraztsova et~al\mbox{.}}{2016a}]%
        {OLPRR16}
\bibfield{author}{\bibinfo{person}{S. Obraztsova}, \bibinfo{person}{O. Lev},
  \bibinfo{person}{M. Polukarov}, \bibinfo{person}{Z. Rabinovich}, {and}
  \bibinfo{person}{J.~S. Rosenschein}.} \bibinfo{year}{2016}\natexlab{a}.
\newblock \showarticletitle{Non-myopic voting dynamics: An optimistic
  approach}.
\newblock In \bibinfo{booktitle}{{\em Proc. of the 10th Multidisciplinary
  Workshop on Advances in Preference Handling (M-PREF)}}.
\newblock


\bibitem[\protect\citeauthoryear{Obraztsova, Rabinovich, Elkind, Polukarov, and
  Jennings}{Obraztsova et~al\mbox{.}}{2016b}]%
        {obraztsova2016trembling}
\bibfield{author}{\bibinfo{person}{Svetlana Obraztsova},
  \bibinfo{person}{Zinovi Rabinovich}, \bibinfo{person}{Edith Elkind},
  \bibinfo{person}{Maria Polukarov}, {and} \bibinfo{person}{Nicholas~R
  Jennings}.} \bibinfo{year}{2016}\natexlab{b}.
\newblock \showarticletitle{Trembling hand equilibria of plurality voting}.
  AAAI Press/International Joint Conferences on Artificial Intelligence.
\newblock


\bibitem[\protect\citeauthoryear{Osborne and Rubinstein}{Osborne and
  Rubinstein}{2003}]%
        {osborne2003sampling}
\bibfield{author}{\bibinfo{person}{Martin~J Osborne} {and}
  \bibinfo{person}{Ariel Rubinstein}.} \bibinfo{year}{2003}\natexlab{}.
\newblock \showarticletitle{Sampling equilibrium, with an application to
  strategic voting}.
\newblock \bibinfo{journal}{{\em Games and Economic Behavior\/}}
  \bibinfo{volume}{45}, \bibinfo{number}{2} (\bibinfo{year}{2003}),
  \bibinfo{pages}{434--441}.
\newblock


\bibitem[\protect\citeauthoryear{Regenwetter, Ho, and Tsetlin}{Regenwetter
  et~al\mbox{.}}{2007}]%
        {regenwetter2007sophisticated}
\bibfield{author}{\bibinfo{person}{Michel Regenwetter},
  \bibinfo{person}{Moon-Ho~R Ho}, {and} \bibinfo{person}{Ilia Tsetlin}.}
  \bibinfo{year}{2007}\natexlab{}.
\newblock \showarticletitle{Sophisticated approval voting, ignorance priors,
  and plurality heuristics: A behavioral social choice analysis in a
  Thurstonian framework.}
\newblock \bibinfo{journal}{{\em Psychological Review\/}}
  \bibinfo{volume}{114}, \bibinfo{number}{4} (\bibinfo{year}{2007}),
  \bibinfo{pages}{994}.
\newblock


\bibitem[\protect\citeauthoryear{Reijngoud and Endriss}{Reijngoud and
  Endriss}{2012}]%
        {RE12}
\bibfield{author}{\bibinfo{person}{Annemieke Reijngoud} {and}
  \bibinfo{person}{Ulle Endriss}.} \bibinfo{year}{2012}\natexlab{}.
\newblock \showarticletitle{Voter response to iterated poll information}. In
  \bibinfo{booktitle}{{\em 11th}}. \bibinfo{pages}{635--644}.
\newblock


\bibitem[\protect\citeauthoryear{Riker and Ordeshook}{Riker and
  Ordeshook}{1968}]%
        {riker1968theory}
\bibfield{author}{\bibinfo{person}{William~H Riker} {and}
  \bibinfo{person}{Peter~C Ordeshook}.} \bibinfo{year}{1968}\natexlab{}.
\newblock \showarticletitle{A Theory of the Calculus of Voting}.
\newblock \bibinfo{journal}{{\em American political science review\/}}
  \bibinfo{volume}{62}, \bibinfo{number}{1} (\bibinfo{year}{1968}),
  \bibinfo{pages}{25--42}.
\newblock


\bibitem[\protect\citeauthoryear{Tal, Meir, and Gal}{Tal et~al\mbox{.}}{2015}]%
        {TalMG15}
\bibfield{author}{\bibinfo{person}{Maor Tal}, \bibinfo{person}{Reshef Meir},
  {and} \bibinfo{person}{Ya'akov~(Kobi) Gal}.} \bibinfo{year}{2015}\natexlab{}.
\newblock \showarticletitle{A Study of Human Behavior in Online Voting}. In
  \bibinfo{booktitle}{{\em Proceedings of the 2015 International Conference on
  Autonomous Agents and Multiagent Systems, {AAMAS} 2015, Istanbul, Turkey, May
  4-8, 2015}}. \bibinfo{pages}{665--673}.
\newblock
\newblock
\shownote{Full version available from \url{https://tinyurl.com/yczxugoj}.}


\bibitem[\protect\citeauthoryear{Tversky and Kahneman}{Tversky and
  Kahneman}{1974}]%
        {tversky1974judgment}
\bibfield{author}{\bibinfo{person}{Amos Tversky} {and} \bibinfo{person}{Daniel
  Kahneman}.} \bibinfo{year}{1974}\natexlab{}.
\newblock \showarticletitle{Judgment under uncertainty: Heuristics and biases}.
\newblock \bibinfo{journal}{{\em science\/}} \bibinfo{volume}{185},
  \bibinfo{number}{4157} (\bibinfo{year}{1974}), \bibinfo{pages}{1124--1131}.
\newblock


\bibitem[\protect\citeauthoryear{Tyszler and Schram}{Tyszler and
  Schram}{2016}]%
        {tyszler2016information}
\bibfield{author}{\bibinfo{person}{Marcelo Tyszler} {and}
  \bibinfo{person}{Arthur Schram}.} \bibinfo{year}{2016}\natexlab{}.
\newblock \showarticletitle{Information and strategic voting}.
\newblock \bibinfo{journal}{{\em Experimental economics\/}}
  \bibinfo{volume}{19}, \bibinfo{number}{2} (\bibinfo{year}{2016}),
  \bibinfo{pages}{360--381}.
\newblock


\bibitem[\protect\citeauthoryear{Van~der Straeten, Laslier, and Blais}{Van~der
  Straeten et~al\mbox{.}}{2013}]%
        {van2013vote}
\bibfield{author}{\bibinfo{person}{Karine Van~der Straeten},
  \bibinfo{person}{Jean-Fran{\c{c}}ois Laslier}, {and}
  \bibinfo{person}{Andr{\'e} Blais}.} \bibinfo{year}{2013}\natexlab{}.
\newblock \showarticletitle{Vote au Pluriel: How people vote when offered to
  vote under different rules}.
\newblock \bibinfo{journal}{{\em PS: Political Science \& Politics\/}}
  \bibinfo{volume}{46}, \bibinfo{number}{2} (\bibinfo{year}{2013}),
  \bibinfo{pages}{324--328}.
\newblock


\bibitem[\protect\citeauthoryear{Van~der Straeten, Laslier, Sauger, and
  Blais}{Van~der Straeten et~al\mbox{.}}{2010}]%
        {van2010strategic}
\bibfield{author}{\bibinfo{person}{Karine Van~der Straeten},
  \bibinfo{person}{Jean-Fran{\c{c}}ois Laslier}, \bibinfo{person}{Nicolas
  Sauger}, {and} \bibinfo{person}{Andr{\'e} Blais}.}
  \bibinfo{year}{2010}\natexlab{}.
\newblock \showarticletitle{Strategic, sincere, and heuristic voting under four
  election rules: an experimental study}.
\newblock \bibinfo{journal}{{\em Social Choice and Welfare\/}}
  \bibinfo{volume}{35}, \bibinfo{number}{3} (\bibinfo{year}{2010}),
  \bibinfo{pages}{435--472}.
\newblock


\bibitem[\protect\citeauthoryear{Wright and Leyton-Brown}{Wright and
  Leyton-Brown}{2010}]%
        {wright2010beyond}
\bibfield{author}{\bibinfo{person}{James~R Wright} {and} \bibinfo{person}{Kevin
  Leyton-Brown}.} \bibinfo{year}{2010}\natexlab{}.
\newblock \showarticletitle{Beyond equilibrium: predicting human behaviour in
  normal form games}. In \bibinfo{booktitle}{{\em AAAI}}.
\newblock


\bibitem[\protect\citeauthoryear{Yosef, Naamani-Dery, Obraztsova, Rabinovich,
  and Bannikova}{Yosef et~al\mbox{.}}{2017}]%
        {yosef2017haste}
\bibfield{author}{\bibinfo{person}{David~Ben Yosef}, \bibinfo{person}{Lihi
  Naamani-Dery}, \bibinfo{person}{Svetlana Obraztsova}, \bibinfo{person}{Zinovi
  Rabinovich}, {and} \bibinfo{person}{Marina Bannikova}.}
  \bibinfo{year}{2017}\natexlab{}.
\newblock \showarticletitle{Haste makes waste: a case to favour voting bots}.
  In \bibinfo{booktitle}{{\em Proceedings of the International Conference on
  Web Intelligence}}. ACM, \bibinfo{pages}{419--425}.
\newblock


\end{thebibliography}
\end{document}